\documentclass{egpubl}
\usepackage{eurovis2026}

\EuroVisShort  

\usepackage[T1]{fontenc}
\usepackage{dfadobe}

\usepackage{cite}  
\BibtexOrBiblatex
\electronicVersion
\PrintedOrElectronic
\ifpdf \usepackage[pdftex]{graphicx} \pdfcompresslevel=9
\else \usepackage[dvips]{graphicx} \fi

\usepackage{egweblnk}
\usepackage{xcolor}

\newcommand{\redash}[1]{\textsc{ReDash}{#1}}

\newcommand{\gustavo}[1]{{\color{brown} \noindent Gustavo: [{#1}]}}

\newcommand{\rr}[1]{{\color{black} {#1}}}

\newcommand{\qte}[2]{\textit{``#1''}~($P{#2}$)}

\newcommand{\bundle}[1]{\texttt{\textbf{#1}}}

\title{Once Again, with Style: Understanding and Supporting \\ Partial Reuse in Dashboard Authoring}

\author[N. Sultanum, G. Moreira, \& A. Srinivasan]
{\parbox{\textwidth}{\centering 
        Nicole Sultanum$^{1}$\orcid{0000-0001-8608-1427}
        Gustavo Moreira$^{2}$\orcid{0000-0002-4762-7703}
        Arjun Srinivasan$^{1}$\orcid{0000-0001-8901-1256}
        }
        \\
{\parbox{\textwidth}{\centering $^1$Tableau Research, Seattle, USA\\
         $^2$University of Illinois, Chicago, USA
       }
}
}

\begin{document}


\maketitle
\begin{abstract}
 Presentation-oriented tasks including formatting and layout design are critical but often neglected aspects of dashboard authoring given their labor intensive nature. In this work, we follow a user-centered design approach to explore ways that \textit{partial} \textit{reuse} of pre-existing dashboards may support the dashboard design process. Based on collective feedback from 10 professional dashboard creators, we contribute: (a) findings from a formative study characterizing dashboard reuse needs and challenges; and (b) reflections and opportunities from a concept validation study with \redash{,} a design probe for partial reuse of \rr{dashboard presentation features (style and layout) from multiple sources}.


\end{abstract}  
\raggedbottom

\section{Introduction and Background}
Dashboard authoring is a time-consuming task requiring a broad range of skills and multidisciplinary teams~\cite{ma2020ladv}. Within that, presentation-oriented tasks such as \textit{styling and formatting} are notably ill supported in commercial dashboard platforms~\cite{bach2022dashboard, tory2021finding}, requiring time-intensive fine tuning and rote replication of work already done for previous dashboards.  

Prior work on supporting dashboard presentation tasks has explored different workflows including template generation from reference images and sketches~\cite{ma2020ladv} and natural language prompting~\cite{shen2025dashchat}. Automated techniques have also been explored to leverage design rules~\cite{lin2023dminer} and domain-specific languages~\cite{jiang2022mod2dash} to model design choices~\cite{lin2023dminer, chen2020composition, shi2022colorcook}, recommend effective layouts~\cite{chen2020composition, zeng2023semi}, and configure new dashboards~\cite{shen2025dashchat,ma2020ladv}.
While helpful for rapid authoring, these approaches often result in an ``all or nothing" experience, requiring users to accept the suggested styles and formats as-is or fine-tune individual components, resulting in a tedious process comparable to manual authoring.

Inspired by the notion of  ``mashup tools'' alluded by Tory et al.~\cite{tory2021finding}  where ``a dashboard user could easily snag content from different dashboards'', we believe there is an untapped opportunity to make dashboard presentation tasks like styling and formatting easier by leveraging the idea of \textit{dashboard reuse}. The notion of reusing artifacts has been explored broadly in web and UX design~\cite{fitzgerald2008copystyler, kumar2011bricolage, lu2025misty} as well as for authoring individual charts~\cite{jung2017chartsense, mendez2016ivolver, harper2017converting,bako2022understanding} but has received relatively \rr{little} attention in the context of dashboard authoring.

In this work, we fill this gap via a user-centered approach to characterize the notion of \textit{reuse} for dashboard authoring. Through a formative study with professional dashboard creators, we elicited current reuse practices, challenges, and needs~(\S\ref{sec:formative}). We then developed  
\redash{,} a proof-of-concept to facilitate \rr{\textit{partial reuse of dashboard presentation features} (layout and styling) \textit{from multiple  sources}} (\S\ref{sec:redash})\rr{, created as a \rr{design probe} to help refine our understanding of}
dashboard reuse \rr{opportunities} (\S\ref{sec:validation}), and inform meaningful research directions for future work (\S\ref{sec:reflections}).

\vspace{-0.5em}

\section{Formative Study}
\label{sec:formative}

Given the limited knowledge around dashboard reuse practices, 
\rr{we conducted}
a formative study with
7 professional dashboard creators (P1-P7), recruited \rr{from online communities}. Occupations spanned BI and visualization engineers (2), data and BI leaders (managers, directors) (4), and a consultant (1) averaging 8.5 years ($min=5$, $max=15$) of self-reported experience in data related fields. Sessions were 1-hour long, encompassing a semi-structured interview on current practices and challenges, and a hands-on reuse exercise on a digital whiteboard. After choosing one of two scenarios \rr{they were most comfortable with (between business and infographics)}, participants were asked to create a dashboard \textit{concept design} while ``borrowing inspiration'' from a curated collection of 9-11 dashboards references, documenting it clearly enough to ``hand it off to someone else to implement their vision'' (\autoref{fig:concept-design}); the full study plan is provided in supplemental materials (Appendix A).
Sessions were screen- and audio- recorded and transcribed, followed by a thematic analysis of transcripts and documented \textit{reuse instances} in dashboard concept designs (34 total, across 7 participants). 




\begin{figure}
    \centering
    \includegraphics[width=\linewidth]{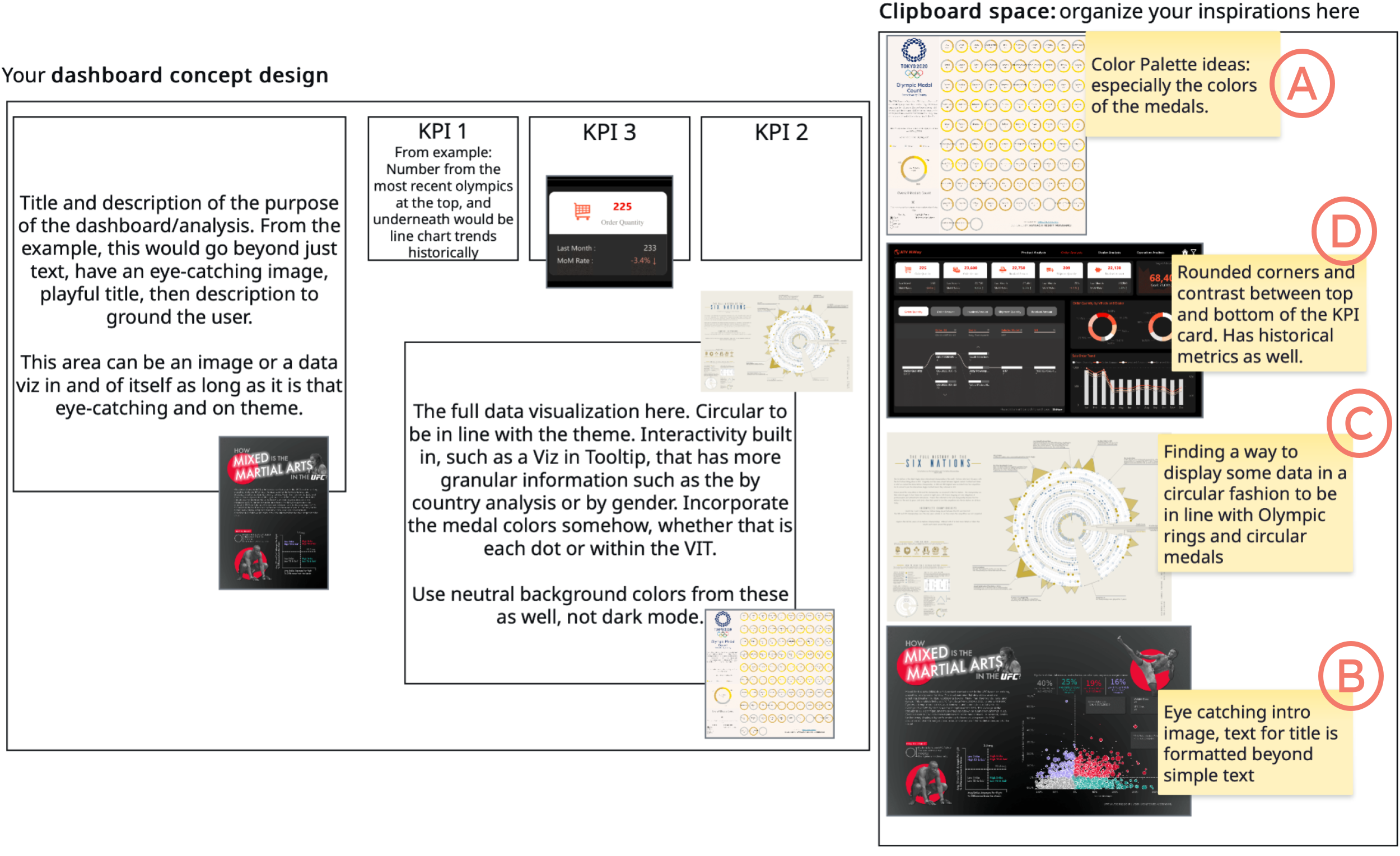}
    \caption{A concept design from a study participant, featuring reuse instances documented as post-it notes, text/shape annotations, and dashboard  cut-outs. Refer to \S\ref{sec:formative-practices} for (A) - (D) callouts.
    \vspace{-1em}
    }
    \label{fig:concept-design}
    
\end{figure}

\subsection{The Why: Challenges to Dashboard Authoring}
\label{sec:formative-challenges}

\noindent\textbf{(Ch1) Redundancy.}
Some (3/7 participants) commented on the repetitive nature of authoring work, in that there are \qte{no new problems we're solving that haven't already been solved within our existing suite of dashboards}{2}, and that \qte{people need very slight variations of the same thing}{5}.
However, options to leverage existing components and styling are often limited to ``cracking'' dashboards open, and that \qte{the [reuse] functionality doesn't exist in a way that saves enough time to bother}{4}. 



\noindent \textbf{(Ch2) Styling and Layout.}
Unlike data pipelines powered by centralized and reusable workflows,
aspects of dashboard \textit{presentation} such as fill color, outlines, padding and text formatting (i.e.,  layout and styling), emerged as particularly time-consuming and under-supported (7/7) while also critical in many business settings where \qte{things need to be pixel perfect}{7} and \qte{dashboards that make their directors eyes’ shine are key}{1}.

\noindent \textbf{(Ch3) Consistency.}
Many (4/7) underscored a need for dashboard portfolios to \qte{be consistent and to have a consistent theme}{7}.
This becomes challenging as scale grows and larger teams are involved, which in turn calls for formal documentation standards such as style guides~\cite{onlinestyleguides, sultanum2024instruction} to keep everyone aligned. Enforcing them, however, is also a largely manual task, as not everything can be systematically controlled: \qte{it's very hard, with the size of our book of work, to look at each work and every definition}{2}.
Dashboard templates~\cite{ma2020ladv} are a popular alternative to enforce consistency (6/7), but many (4/7) consider them over-restricting: \rr{\qte{I want my team to 
design for the purpose of the analysis, not just
fit in well [to a layout]. 
}{5}.
}

\subsection{The How: Opportunities for Dashboard Reuse}
\label{sec:formative-practices}



\noindent\textbf{(Op1) Support flexible scoping for partial reuse.}
A common thread to all three challenges \textbf{(Ch1-3)} is a need for more flexible forms of \textit{partial} reuse. All  (7/7) voiced a desire for better tools to \qte{copy and paste chunks of things}{4}, and replicating \qte{well-designed pieces}{6} as component bundles (5/7) featuring specific combinations of attributes such as style, layout, data transformations, and interactive behavior (e.g., BANs and filter panel widgets).
Reuse intents from concept designs also heavily leaned towards bundled components (\textasciitilde73\% of instances), underscoring a need to support various source/target scope selections.

\noindent\textbf{(Op2) Support reuse intents at various levels of granularity.}
A closer look on reuse instances in concept designs reveals a broad range of reuse intents, from specific and clearly computable reuse needs (e.g., reusing a \textit{``color palette''},  \autoref{fig:concept-design}A) to more abstract, inspiration-level intents with unclear execution (e.g.,\textit{``eye-catching intro image''}, \autoref{fig:concept-design}B). This suggests a need for a computational approach that  accommodates various degrees of intent granularity
and is able to ``fill in the blanks'' of under-specified intents~\cite{Lundgard2021AccessibleVV}.

\noindent\textbf{(Op3) Facilitate reuse of styling and layout.}
\rr{\textit{Styling} (7/7) and \textit{layout} (5/7) reuse were among the most cited needs. In} concept designs, \textasciitilde50\% of reuse instances referred to layout inspirations (encoded as placeholders or nominally referred in annotations (e.g., \textit{``display data in circular fashion''}, \autoref{fig:concept-design}C), and \textasciitilde32\% to styling (e.g., \textit{``rounded corners and contrast between top and bottom''}, \autoref{fig:concept-design}D). 
This denotes dashboard presentation attributes as both a reuse challenge \textbf{(Ch2)} and a priority to support.

\noindent\textbf{(Op4) Reuse from multiple sources.}
\label{sec:formative-practices-sources}
All concept designs featured \rr{elements} from at least 3 different dashboards, highlighting the limitations of one-size-fits-all template approaches\textbf{ (Ch3)} and suggesting a need \rr{for} multiple reuse sources. \rr{Some (3/7)} also mentioned explicitly tracking dashboard inspirations, by bookmarking ``favorites'' on Tableau Public or curating mood boards, emphasizing the role that prior references play in influencing design while underscoring a need to leverage these sources at authoring stages.

\begin{figure*}
  \centering
  \includegraphics[width=\textwidth]{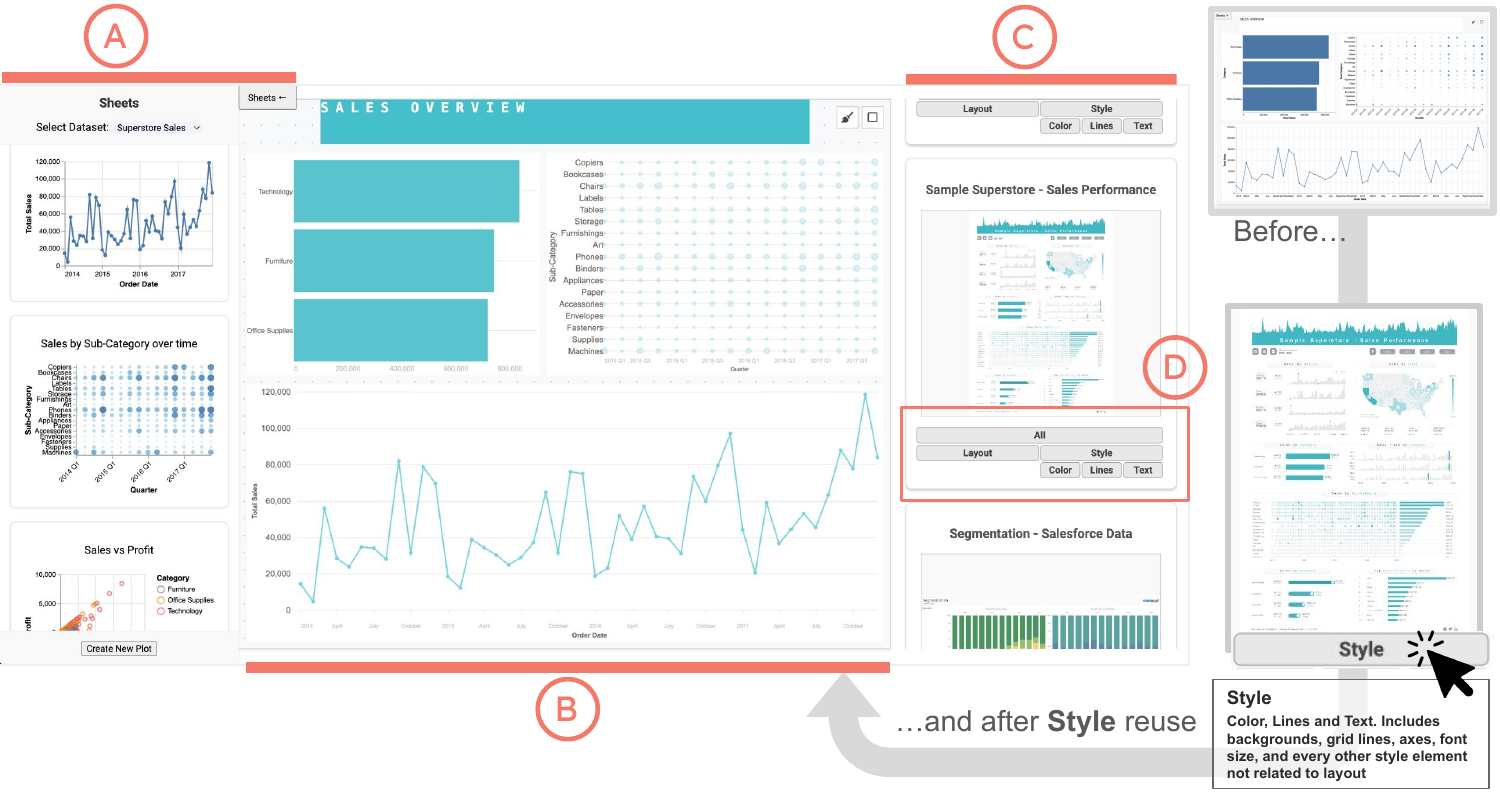}
  \caption{
 The \redash{}  interface, our design probe, featuring: (A) a list of \textit{data-bound components}; (B) the \textit{authoring canvas};
 (C) a \textit{list of dashboard references} available for reuse, and corresponding (D) \textit{design bundles} for quick style and layout reuse. 
\vspace{-1em}
  }
  \label{fig:interface}
\end{figure*}
\vspace{-0.5em}
\section{\rr{The \redash{} design probe}}
\label{sec:redash}

\rr{To refine our understanding of reuse needs, we implemented \redash{}, a scoped but end-to-end dashboard reuse tool to help mediate conversations with users. 
It features \textit{partial} reuse~\textbf{(Op1)} of dashboard \textit{layout} and \textit{styling}~\textbf{(Op3)} from \textit{multiple sources}~\textbf{(Op4)} and at various \textit{levels of granularity}~\textbf{(Op2)}.}

\rr{To streamline focus on presentation features~\textbf{(Op3)}, \redash{} provides a list of  pre-authored \textit{data-bound components} (\autoref{fig:interface}A) that can be dragged onto a dashboard authoring canvas (\autoref{fig:interface}B). A list of \textit{dashboard references} (\autoref{fig:interface}C)  provides multiple sources of reuse \textbf{(Op4)}, each with a multi-level set of \textit{design bundles} (\autoref{fig:interface}D) featuring subsets of presentation features \textbf{(Op1)} at various levels of granularity \textbf{(Op2)}:  
(a) \bundle{Color} (mark, background, and shadow colors), 
(b) \bundle{Lines} (border, grid, axis, and tick formatting),
(c) \bundle{Text} (font family, size, and weight), 
 (d) \bundle{Layout} (relative size, position and spacing); 
 (e) \bundle{Style} (\texttt{Color}, \texttt{Lines}, and \texttt{Text}), and
   (f) \bundle{All} (\texttt{Style} and \texttt{Layout}).
The list of presentation features in each design bundle is visible via tooltips, and can be applied with a single click.}

When a design bundle is applied, \redash{} adaptively matches components between the dashboard reference (\textit{sources}) and the authoring canvas (\textit{targets}), and merges attributes accordingly. This easy, one-click action attempts to ``fill-in-the-blanks'' even when perfect matches aren't available -- e.g., borrowing \texttt{Style} from a bar chart onto a line chart, or creating placeholders for missing components when borrowing \texttt{Layout} \textbf{(Op2)}. 
\rr{But for more controlled reuse operations, users can also select specific source and target components \textbf{(Op1)}. 
This flexibility supports multiple reuse workflows: from one-step operations such as creating placeholders onto a blank canvas from a single layout reuse step, to iterative refinement via scoped reuse operations.}
A video demo and use case scenario (Appendix B) are provided in supplemental materials.



\vspace{-0.5em}
\subsection{Implementation}
\label{sec:redash-implementation}
\redash{} is implemented as a React app with a Flask server, featuring a reuse pipeline \rr{with} two modules. At pre-processing, a \textbf{design extraction} module leverages a multimodal large language model (LLM) to extract dashboard components and computes corresponding \textit{style specs} for dashboard references.
Style specs encompass the list of reusable style and layout attributes for dashboard components that serve as \textit{shared} representation between source and target \rr{selections}, encoded as Vega-Lite~\cite{satyanarayan2017vega} attributes for charts and CSS attributes for all other components.
During authoring, the \textbf{design transfer} module matches source and target component selections \rr{based on their type and size, then modifies style specs by combining (a) an \textit{LLM-merged spec} of matched source-target pairs, with (b) a \textit{programmatic, frequency-based, ``representative'' spec} of all source selections. We found this setup helped reduce LLM-based inconsistencies across runs}. Full implementation details are provided in supplemental materials (Appendix C).

\vspace{-0.5em}
\section{Design Probe Concept Validation }
\label{sec:validation}

To validate the utility of partial reuse and further elicit reuse opportunities, we conducted open-ended feedback sessions with 6 professional dashboard creators (P2, P4, P5, plus new recruits P8-P10) using \redash{} as a design probe~\cite{hutchinson2003technology}. Participant occupations spanned BI \& visualization engineers (3), and data \& BI leaders (managers and directors, 3), and averaged 8.8 years ($min=5$, $max=12$) of self reported experience in data related fields.



Sessions were approximately 
30 min long, and included \rr{(a) brief demonstrations of the tool with usage examples; (b) an interactive think-aloud segment where participants were asked to imagine a dashboard reuse scenario leveraging the provided dashboard references while freely experimenting with \redash{}; and (c) a follow up discussion on their experience with the tool and ideas for improvement.}
\rr{We made small usability improvements between sessions, including tweaks to interactive selection, design transfer, and \textit{look \& feel}.}
Sessions were screen recorded, audio recorded, and transcribed for thematic analysis. Study instruments are provided in supplemental materials (Appendix D).

\vspace{-0.5em}
\subsection{Feedback and (More) Opportunities}

Participants responded positively to the reuse experience \rr{mediated by} \redash{.} \rr{They quickly learned the controls, using the tooltips to understand reuse bundles while experimenting with different bundles.
They appreciated the simplicity,}
saying 
it \qte{makes a ton of sense and feels intuitive}{4} 
and requires \qte{just a couple of clicks}{10}. 
Others valued efficiency, arguing that \redash{} is \qte{definitely is a timesaver}{9}\rr{ and provides a \qte{cool way to speed up the design process}{5}.
Some also \rr{saw potential} in bridging the expertise gap for non-designers ($P2$) and saving time \qte{upskilling junior analysts who don't know how to do a layout}{10}.}

Beyond confirming the value of partial reuse for dashboard authoring,
\rr{user feedback} informed additional design opportunities besides those identified in the formative study (\textbf{Op1}-\textbf{Op4}).

\noindent\textbf{(Op5) Supporting more diverse workflows.}
Participants identified alternative authoring workflows to consider, such as 
 ``filling-in-the-gaps'' within the confines of a partial template: \qte{With AI and templating into the mix, you've got that potential to go, `what else would be interesting', `what would you suggest' `\rr{how} would you lay this out'\,}{10}; and shifting from a data-first to a style-first approach by copying components first and changing the data bindings after \rr{(P2, P4).}
 We see a need to expand modes of partial reuse to support a broader range of authoring workflows.


\noindent\textbf{(Op6) Surfacing style specs for fine-tuning.}
While participants quickly learned to use the design bundles in \redash{}, they craved more fine-grained control over design attributes, from
choosing properties to \qte{keep locked in style}{4} and changes to propagate widely, e.g., 
\qte{remove grid lines [from] all my charts at once}{9}. 
Surfacing style specs to end users (e.g., as design attribute widgets~\cite{swearngin2020scout}) 
 could help them \qte{specify what is sacrosanct}{9}, and potentially manage consistency at scale, although also a design challenge in
 balancing efficiency and intuitiveness.

\noindent\textbf{(Op7) Managing and crediting dashboard references.}
\label{sec:validation-findings-attribution}
Participants suggested tools to manage the list of reusable dashboard references and dashboard components, such as bookmarking (P2, P10) and categorizing as a \qte{catalog of things that says `I like these pieces'\,}{2}. 
They also touched on the topic of \textit{attribution}, underscoring the importance of crediting authors for their work (P4, P5) to help curb plagiarism, 
 while also acknowledging the opportunity to visibly reward contributors for their public service: \qte{like `colors inspired by Kevin' or `layout inspired by Jennifer'\,}{4}.

\section{Future Research}
\label{sec:reflections}

We take a step back and reflect on the collective findings of our user-centered design process to outline promising \textit{research opportunities} in the space of partial dashboard reuse.


\noindent\textbf{Towards a design space for dashboard reuse intent.} 
Formative findings alluded to a range of reuse goals and granularity which informed the design bundles in \redash. 
Building on prior work on dashboard categorizations~\cite{bach2022dashboard, srinivasan2024dashboard, sarikaya2018we} and the reuse of data visualization examples~\cite{bako2022understanding},
we see potential to extend our preliminary observations to formalize a design space of dashboard reuse mechanics, which could support the development of higher-order reuse intelligence. 
Feature-wise, while \redash{} implements a meaningful subset of reuse features for layout and style reuse  (including chart reuse), the viable scope of partial dashboard reuse is much broader. Reusing aspects of data connections~\cite{snyder2025challenges,harper2017converting,savva2011revision} and interactive behavior~\cite{North2000SnaptogetherVA},  leveraging multi-view relationships for coordinated actions and relative placement~\cite{srinivasan2024dashboard, lin2023dminer}, further enforcing consistency across views~\cite{qu2017keeping,chen2020composition}, and matching semantic tone of text~\cite{sultanum2024instruction, lisnic2025plume} are but a few examples.

\noindent\textbf{Augmenting diverse authoring workflows with reuse.}
The diversity of authoring workflows calls for a multi-pronged reuse approach. While \redash{} supports redefining styles for data-bound elements, there is potential to mediate templated dashboard and components~\cite{ma2020ladv} via lazy data-binding~\cite{liu2018data}. Similar to the whiteboard exercise (\S\ref{sec:formative}), we envision the utility of in-canvas annotations in documenting fine-grained reuse intents for AI interpretation that could help mitigate the interaction and deictic challenges of designing with pure chat-based interfaces~\cite{shen2025dashchat}. 


\noindent\textbf{Bridging dashboard reuse and best practices.}
We argue that documented dashboard design principles can play a role in scaffolding dashboard reuse. 
Incorporating style guides~\cite{onlinestyleguides, sultanum2024instruction} and best practices~\cite{moritz2018formalizing, setlur2023heuristics} into the design transfer could help improve reuse outcomes and potentially scaffold novice designers to follow dashboard design and communication standards \cite{wexler2017big}.
Conversely, we also see an opportunity for dashboard reuse features to support inference of best practices by generalizing design rules~\cite{lin2023dminer} from frequent reuse actions, which could be useful not only for wider dissemination of best practices but also in organically capturing personal design styles~\cite{Kim2019DataSelfieEP}.

\noindent\textbf{Consistency at scale.}
Following remarks on attribution, the manual labor involved in maintaining consistency, and the often collaborative nature of professional dashboard authoring practices~\cite{ma2020ladv}, we argue that source-tracking mechanisms may be useful to streamline consistency at scale via automatic consistency checks~\cite{qu2017keeping}. Tools to visualize inconsistencies across a dashboard portfolio and selectively propagate changes could be very useful, perhaps again by leveraging design rules~\cite{lin2023dminer, moritz2018formalizing} as programmatic boolean checks~\cite{sultanum2024data}.

\noindent\textbf{Reference recommendation and management.}
Following prior work on the use of data visualizations examples~\cite{bako2022understanding}, our findings point to dashboard references being often curated but not easily revisited. 
Ideas around web design example galleries~\cite{lee2010designing} align with participant feedback and could be well suited to dashboard references. 
Dashboard-specific techniques for reference discovery and management would be worth investigating further; plus features to surface new references to a user's catalog, organize references into meaningful groups, and recommend references relevant to ongoing authoring tasks.


\section{Conclusion}
\label{sec:conclusion}

We introduce the idea of \textit{partial dashboard reuse} as an approach to aid dashboard authoring by helping authors leverage elements from pre-existing dashboards.
We distill this idea through a user-centered design process involving formative studies with dashboard authors, a proof-of-concept interface \redash{,} and feedback sessions, which informed several opportunities for further \rr{research}.

Limitations of our work include the small participant pool ($n=10$, although consisting of established experts) and the confined scope of \redash{} as a design probe \rr{(e.g., limited} editing features and LLM safeguards).
We also focus on the dashboard author's perspective, precluding in-the-wild assessments and perspectives from dashboard consumers.
However, the formative studies and feedback on \redash{} as a design probe highlight promise in the idea of dashboard reuse as a concept and offer important lessons for future work. Our work points to a clear need for better dashboard reuse support,
\rr{revealing opportunities for} improved interaction mechanics and user control, \rr{while introspecting} on associated challenges such as ethical \rr{and at-scale }reuse.



\vspace{-0.7em}

\bibliographystyle{eg-alpha-doi} 
\bibliography{references}       

\end{document}